\documentclass{aa}
\usepackage{graphicx}
\usepackage{natbib}

\newcommand{\Alf}{Alfv\'{e}n\,}
\newcommand{\Alfvenic}{Alfv\'{e}nic\,}
\begin{document}
%
\title{Application of a MHD hybrid solar wind model with latitudinal dependences to \textbf{{\sc Ulysses}} data at minimum}
\titlerunning{Application of a MHD hybrid solar wind model to \textbf{{\sc Ulysses}} data}
\authorrunning{Aib\'{e}o \emph{et al}}

\author{A. Aib\'{e}o
          \inst{1,2},
          J.J.G. Lima
          \inst{1,3}
          \and
            C. Sauty
            \inst{4}
           }

\institute{Centro de Astrof\'{\i}sica da Universidade do Porto,
Rua das Estrelas, 4150-762 Porto, Portugal \and Departamento de
Engenharia Mec\^{a}nica e Gest\~{a}o industrial da Escola Superior
de Tecnologia de Viseu, Campus Polit\'{e}cnico de Viseu, 4105
Viseu, Portugal \and Departamento de Matem\'{a}tica Aplicada da
Faculdade de Ci\^{e}ncias, Universidade do Porto, Rua do Campo
Alegre, 657, 4169-007 Porto, Portugal \and Observatoire de Paris,
L.U.Th., 92190 Meudon, France}

\offprints{ Alexandre Aib\'{e}o, \email{aaibeo@demgi.estv.ipv.pt}}
\abstract
{}
 {In a previous work, \textsc{Ulysses} data was analyzed to
build a complete axisymmetric MHD solution for the solar wind at
minimum including rotation and the initial flaring of the solar
wind in the low corona. This model has some problems in
reproducing the values of magnetic field at 1 AU despite the
correct values of the velocity. Here, we intend to extend the
previous analysis to another type of solutions and to improve our
modelling of the wind from the solar surface to 1 AU.}
{We compare the previous results to those obtained with a fully
helicoidal model and construct a hybrid model combining both
previous solutions, keeping the flexibility of the parent models
in the appropriate domain. From the solar surface to the \Alf\,
point, a three component solution for velocity and magnetic field
is used, reproducing the complex wind geometry and the well-known
flaring of the field lines observed in coronal holes. From the
Alfv\'{e}n radius to 1 AU and further, the hybrid model keeps the
latitudinal dependences as flexible as possible, in order to deal
with the sharp variations near the equator and we use the
helicoidal solution, turning the poloidal streamlines into radial
ones.} {Despite the absence of the initial flaring, the helicoidal
model and the first hybrid solution suffer from the same low
values of the magnetic field at 1 AU. However, by adjusting the
parameters with a second hybrid solution, we are able to reproduce
both the velocity and magnetic profiles observed by
\textsc{Ulysses} and a reasonable description of the low corona,
provided that a certain amount of energy deposit exists along the
flow.}
{The present paper shows that analytical axisymmetric solutions
can be constructed to reproduce the solar structure and dynamics
from 1 solar radius up to 1 AU.}

\keywords{MHD - solar wind - sun - plasmas}
\date{Received ****** **, 2006; accepted *** **, 2006}
\maketitle

\section{Introduction}
Since \citeauthor{Parker58}'s model (1958), many studies have been
presented to explain and predict the features and properties of
the solar wind, mainly following two different, yet complementary,
approaches, kinetic and fluid approximations.

These techniques are able to reproduce certain aspects of the
observed solar wind but both show some limitations, mainly due to
the complexity of the several acceleration mechanisms, the
uncertainties concerning the origin of the fast solar wind, the
associated problem of coronal heating, etc. Different models have
been presented improving results of the acceleration. Two fluid
models
and more recently three-fluid models, (e.g. \citeauthor{Ofman00},
2000; \citeauthor{Zouganelisetal04}, 2004) have been constructed
to explore the kinetic aspects of the wind acceleration by
supra-thermal electrons in the collisionless region far from the
Sun. All models still have difficulties avoiding very high
temperature for the electrons. Other sources of heating such as
turbulence \citep{LandiPantellini03}
 or Alfv\'en waves \citep{UsmanovGoldstein03,Grappinetal02}
may also explain the acceleration by lowering the effective
polytropic index of the flow. This point is not yet resolved and
we shall not address this question here. Instead we will invoke
the need for turbulence or Alfv\'en wave damping in our solutions.

Another approach consists of constructing MHD solutions to analyze
the 3D structure of the wind, almost independently of the heating
source. Various models have been constructed, either 2-D ones able
to describe the low corona of the sun up to 10 solar radii (e.g.
\citeauthor{PneumannKopp71}, 1971; \citeauthor{Steinolfsonetal82},
1982; \citeauthor{Cupermanetal90}, 1990 and references therein),
the slow solar wind inside the brightness boundary in coronal
streamers, (e.g. \citeauthor{NerneySuess05}, 2005 and references
therein), 2-D ones for larger distances and models for all range
of distances. Those models were proposed because the flaring of
the streamlines in polytropic winds favors the acceleration.
Recent observations by \citet{WangSheeley03} showed, however, that
this may not be the case for the real solar wind. This favors a
description of the 3D structure of the solar wind using
self-similar MHD analytical solutions for non polytropic winds
\citep{TsinganosSauty92,LPT01}. In the first of these two models
it has been shown that the flaring of the lines may instead limit
the acceleration of the wind.

An increasing amount of observational data is now available. {\sc
Ulysses} measured for the first time the magnetic field, the
dynamics and the temperature of the wind around 1 AU out of the
ecliptic plane \citep{McComasetal00}. Data from  ACE
\citep{Stoneetal98}, WIND \citep{Acunaetal95}, \textsc{SoHO} UVCS
(e.g. \citeauthor{WooHabbal05}, 2005), LASCO (e.g.
\citeauthor{LewisSimnett02}, 2002) and \textsc{SoHO} CDS
\citep{Gallagheretal99} are providing new insights into the origin
of the solar wind within coronal holes. Doppler Scintillation
measurements \citep{WooGazis94} also brought new constraints to
solar wind modelling.

Semi-empirical models that use data to set boundary conditions for
a numerical approach to the problem have also been proposed (e.g.
\citet{Steinolfsonetal82,SittlerGuhathakurta99,Grothetal99}).
Nevertheless some doubts on the boundaries of some simulations are
still present \citep{Vlahakisetal00}. More recently, some new
developments have suggested that numerical simulations can benefit
greatly from an analytical treatment (e.g.
\citeauthor{KeppensGoedbloed00}, 2000;
\citeauthor{UsmanovGoldstein03}, 2003; \citeauthor{Hayashi05},
2005). Numerical simulations are still quite time-consuming
although this is rapidly improving. However, there are other
limitations such as maintaining divergence-free magnetic fields,
limiting the effects of numerical magnetic diffusivity or solving
the 3D structure of the wind including rotation even at large
distances. Note that the main problem with present simulations is
the existence of a numerical magnetic diffusivity (e.g.
\citeauthor{Grappinetal02}, 2002). This is why we propose to
construct semi-analytical models which are less sophisticated than
numerical simulations but simpler to handle and more versatile.
They also provide a complementary approach.

In the present work, that follows closely the work of \citeauthor
{SLIT05} (2005 hereafter SLIT05) we focus on the dynamics of the
protons in the solar wind. We apply known MHD analytical models to
{\sc Ulysses} data at solar minimum and test their advantages and
limitations. We will generate an exact wind solution based on the
model of Lima et al. (2001 hereafter LPT01) and on the data fit
made in SLIT05. In SLIT05 two models (LPT01 and \citeauthor{STT99}
1999 hereafter STT99) were used to fit the data. The final
solution was based on solving the differential equations of the
latter of these two models. Finding a solution that complies with
the constraints given by the data fit and the ones from its
topological features is not easy.

Regarding the limitations of LPT01 wind solution (purely radial,
yet very adaptable to the latitudinal dependences) and the ones of
the similar study presented in SLIT05 (see also STT99;
\citeauthor{STT02}, 2002, hereafter STT02 and \citeauthor{STT04},
2004, hereafter STT04), we take into account the advantages of
both models by creating hybrid solutions. These use the 2.5D
features of the STT04 model to describe the solar wind dynamics
from the solar surface towards the Alfv\'{e}n sphere. From the
Alfv\'{e}n point towards 1 AU and beyond, these hybrid solutions
will use the advantages of the LPT01 model in fitting steep
variations of velocity, density and magnetic field with latitude
and expressing the radial behaviour of the solar wind in this
region. However, we still solve the complete set of MHD equations
in the radial domain and not simply the Bernoulli equation along
the streamlines. Thus, the solution remains consistent everywhere.
We discuss the properties of the solutions thus obtained and
physical grounds for their limitations.

We maintain the criteria used to find a good solution from  the
{\sc Ulysses} data fit and the measured values of the physical
quantities at 1 AU. It will be shown later that, for some sets of
parameters, both the LPT01 model and the first hybrid model show
the same problems mentioned in SLIT05, namely in reproducing the
values of the magnetic field at 1 AU from \textsc{Ulysses}. These
will be solved by a judicious choice of parameters in the second
hybrid model that generates a solution consistent with {\sc
Ulysses} data.

\section{Self-similar MHD outflow models from central rotating objects}
The following two axisymmetric wind models are obtained by
self-consistently solving the full system of ideal MHD equations.
In the present work we use spherical coordinates $[\emph{r},
\theta, \phi]$. All quantities have been normalized to their
values at the Alfv\'{e}n radius along the polar axis, similarly to
SLIT05.  They will be identified by the subscript *, \emph{i.e.}
$V_{*}$, $\rho_{*}$ and $B_{*}$ for velocity, density and magnetic
fields at the Alfv\'{e}n polar point, respectively. All equations
will be presented in a normalized form where the distance to the
solar surface is related to the real distance by $R\equiv r/r_*$.

\subsection{Model A with flaring streamlines}
In model A all three components of the velocity and magnetic
fields are accounted for (STT99, STT02, STT04). Nevertheless, an
expansion up to first order in latitude of the forces is performed
by using harmonics with polar values as references. Such a
procedure makes the whole system analytical tractable and also
describes the helio-latitudinal variations of the wind quantities.
The fields describing the outflow dynamics are
%
\begin{eqnarray}
V_r(R,\theta) & = & V_* \frac{f
M^2}{R^2}\frac{\cos\theta}{\sqrt{1+\delta f \sin^{2}\theta}}
\label{STTVr}\\
V_\theta(R,\theta) & = & -V_*
\frac{M^2}{2R}\frac{df}{dR}\frac{\sin\theta} {\sqrt{1+\delta f
\sin^{2}\theta}}
\label{STTVtheta}\\
V_\phi(R,\theta) & = & \lambda V_* \left
(\frac{1-fM^2/R^2}{1-M^2}\right) \frac{R \sin\theta}
{\sqrt{1+\delta f \sin^{2}\theta}}
\label{STTVphi}\\
B_r(R,\theta) & = & B_* \frac{f}{R^2}\cos\theta
\label{STTBr}\\
B_\theta(R,\theta) & = & -B_* \frac{1}{2R}\frac{df}{dR}\sin\theta
\label{STTBtheta}\\
B_\phi(R,\theta) & = & \lambda B_* \left
(\frac{1-f/R^2}{1-M^2}\right) R\sin\theta
\label{STTBphi}\\
\rho(R,\theta) & = & \frac{\rho_*}{M^2}\left(1+\delta f
\sin^{2}\theta \right)
\label{STTrho}\\
P(R,\theta) & = & \frac{1}{2}\rho_* V_*^2 \left(\Pi\left(1+\kappa
f \sin^2\theta \right)+C\right), \label{STT_model_end}
\end{eqnarray}
where $V_r$, $V_\theta$, $V_\phi$ are the three components of the
velocity field, $B_r$, $B_\theta$, $B_\phi$, the three components
of the magnetic field, $\rho$, the density, $P$, the pressure and
$C$ is a constant. There are three functions of $R$, namely $M$,
$f$ and $\Pi$.
\subsection{Model B with helicoidal/radial streamlines}
Model B assumes a simpler geometry with radial stream and field
lines in the poloidal plane (\emph{i.e.} zero $\theta$ components
of the velocity and magnetic fields). It is more versatile at
reproducing steep latitudinal variations (LPT01). In this case,
the fields describing the outflow dynamics are
\begin{eqnarray}
    \label{LPT_model_begin}
V_r(R,\theta) & = & V_*
\frac{M^2}{R^2}\sqrt{\frac{1+\mu\sin^{2\epsilon}\theta} {1+\delta
\sin^{2 \epsilon}\theta}}
\label{LPTVr}\\
V_{\phi}(R,\theta) & = & \lambda V_*
\left(\frac{1-M^2/R^2}{1-M^2}\right)\frac{R\sin^{\epsilon}\theta}
{\sqrt{1+\delta\sin^{2\epsilon}\theta}}
\label{LPTVtheta}\\
B_r(R,\theta) & = & \frac{B_*}{R^2}\sqrt{{1+\mu
\sin^{2\epsilon}\theta}}
\label{LPTBr}\\
B_\phi(R,\theta) & = & \lambda B_* \left
(\frac{1-1/R^2}{1-M^2}\right) R \sin^{\epsilon} \theta
\label{LPTBphi}\\
\rho(R,\theta) & = &
\frac{\rho_*}{M^2}\left(1+\delta\sin^{2\epsilon}\theta\right)
\label{LPTrho}\\
P(R,\theta) & = & \frac{1}{2}\rho_* V_*^2\left(\Pi_0+\Pi_1
\sin^{2\epsilon}\theta \right), \label{LPT_model_end}
\end{eqnarray}
where the same notation is used and $\Pi_0$ and $\Pi_1$ are
functions of $R$.
\subsection{Geometry of the solutions}
The relevant wind type solutions cross various critical points
related to the non-linearity of the system of equations and its
mixed elliptic/hyperbolic nature (see for instance Tsinganos et
al., \citeyear{Tsinganosetal96}; STT04). Each model is described
by three functions of $R$, $\emph{M(R)}$, $\Pi(R)$ and
$\emph{f(R)}$ for model A,  $\emph{M(R)}$, $\Pi_0(R)$ and
$\Pi_1(R)$ for model B. The function \emph{f(R)} characterizes the
geometry of the fieldlines and expresses the expansion factor. For
a fully radial poloidal fieldline (i.e. an helicoidal pattern of
the lines in 3D) we have $f=1$ which is the case of model B. The
function $M(R)$ describes the poloidal Alfv\'{e}n Mach number
which is unity at the Alfv\'{e}n radius. At this point the kinetic
energy overtakes the magnetic one. A limitation of both models
comes from their self-similar nature. Thus the \Alf Mach number is
independent of latitude and therefore the \Alf iso-surfaces are
spherical. The functions $\Pi$, $\Pi_0$ and $\Pi_1$ are determined
by numerical and analytical techniques that are explained in STT02
and LPT01.

\section{A complete solution with helicoidal\textbf{/radial} streamlines}\label{build_B}
In SLIT05 we fitted \textsc{Ulysses} data using the latitude
dependence of models A and B. We have shown that both models yield
similar parameters. The system of ODEs was integrated exclusively
using model A. Conversely, in this section we use the ODEs of
model B to derive a full solution and compare the results with the
ones from SLIT05. The model flexibility provides a better fit of
the latitudinal functions, which may be crucial in dealing with
the poloidal data at 1 AU. Yet, this solution cannot reproduce the
flaring of the streamlines as they remain radial in the poloidal
plane.

\begin{table}
\centering
  \caption{Parameters obtained by data fitting at 1 AU along the polar
axis using the \textsc{Ulysses} hourly average data, except for
$B_{T,1\rm{AU}}$ which is calculated along the equatorial
plane.}\label{table1}
\begin{center}
\begin{tabular}{c c}
  \hline\\
  Parameters & Model B \\
  \\
  \hline\\
  $\ \kappa {f_{1\rm{AU}}}$ & $\;\;\;\;0.35$ \\
  $\ \delta {f_{1\rm{AU}}}$ & $\;\;\;\;1.95$ \\
  $\ f_{1\rm{AU}} $ & $\;\;\;\;1.00$\\
  $\ \epsilon $ & $\;\;\;\;5.64$ \\
  $\ \mu $ & $-\;0.18$ \\
  \\
  \hline\\
  $V_{1\rm{AU}}$ (km/s) & $\;\;\;775$ \\
  $n_{1\rm{AU}}$ (cm$^{-3}$) & $\;\;\;2.48$ \\
  $B_{1\rm{AU}}$ ($\mu$ G) & $\;\;\;30.4$ \\
  $B_{T,1\rm{AU}}$ ($\mu$ G) & $\;\;\;29.5$ \\
  \\
  \hline
\end{tabular}
\end{center}
\end{table}

\subsection{Method for a solution}
The free parameters of the model ($\epsilon$, $\delta$ and $\mu$),
the polar values of the number density, radial velocity and
magnetic field at 1 AU, $n_{1\rm{AU}}$, $V_{1\rm{AU}}$ and
$B_{1\rm{AU}}$, respectively, and the equatorial toroidal magnetic
field at that same distance, $B_{T,1\rm{AU}}$, have already been
constrained by the \textsc{Ulysses} data fitting procedures used
in SLIT05. The end results are summarized in Table \ref{table1}.
From these constrained parameters the latitude functions are well
defined. The radial dependences of the physical quantities are
determined by integrating the ODEs of the model. From the
knowledge of the values at 1 AU of the polar velocity,
$V_{1\rm{AU}}$, density, $\rho_{1\rm{AU}}$, and magnetic field,
$B_{1\rm{AU}}$, it is possible to infer the value of the
Alfv\'{e}n number at that same distance,

\begin{equation}\label{MA1au}
  M_{1\rm{AU}}^2 =\frac{4\pi\rho_{1\rm{AU}}
V_{1\rm{AU}}^2}{B_{1\rm{AU}}^2}\,.
\end{equation}
Simultaneously using Eq.
(\ref{LPTrho}) we obtain another important reference value,
\begin{equation}\label{ro_star}
  \rho_* =\frac{4\pi\rho_{1\rm{AU}}^2
V_{1\rm{AU}}^2}{B_{1\rm{AU}}^2}\,.
\end{equation}

In the original paper (LPT01) the relations that rule the model
are normalized to the solar surface. The critical solution for the
solar wind is calculated based on three simple criteria. The first
one is $V_r(r/r_{\small{\odot}}=1)/V_\odot=1$ and the other two
are the continuity in the acceleration at the \Alf singularity and
fast magnetosonic separatrix. In the present work we keep the two
last criteria but Eqs. (\ref{LPT_model_begin}) to
(\ref{LPT_model_end}) are normalized to the Alfv\'{e}n radius as
in SLIT05. Thus, regarding the first criterion, the definition of
the solar surface, $r_\odot/r_*\equiv R_\odot$, poses a problem.
In order to calculate it we used the value at which the radial
velocity goes to zero or reaches its minimum value. As we expect
the radial velocity to have a very steep variation at those
distances, the corresponding error at the evaluation of the solar
surface will be very small. The final solution should also
reproduce the measured values obtained by \textsc{Ulysses} at 1 AU
(mainly radial velocity, radial and toroidal magnetic field and
density). The total acceleration between the \Alf radius and 1 AU
can be parameterized by,
\begin{equation}\label{eta}
  \eta =\frac{V_{1\rm{AU}}}{V_*} \,.
\end{equation}
Guessing this parameter, we obtain an initial value of $V_*$ and
$B_*=\sqrt{4\pi \rho_*}V_*$. From Eq.
(\ref{LPTBr})
it is possible to determine the value of 1 AU in \Alf radius units,
\begin{equation}\label{R1au}
  R_{1\rm{AU}}=\frac{1 \textsc{AU}}{r_{*(\textrm{AU})}}=\frac{215
  r_\odot}{r_{*(r_\odot)}}=\sqrt{\frac{B_*}{B_{1\rm{AU}}}}
\,.
\end{equation}
The last equation gives the location of the solar surface since
$R_{1\rm{AU}}=r_{1\rm{AU}}/r_*=215\,R_\odot $. Already having  the
anisotropy parameters, $\epsilon$, $\delta$ and $\mu$, we still
need $\lambda$ and $\nu$. Combining Eqs. (\ref{LPTBr}) and
(\ref{LPTBphi}), assuming that we are at large distances
$(R_{1\rm{AU}}>>1)$ and since the lines are radial, which is very
reasonable at 1 AU, we get, from Eq. (\ref{LPTBphi}) applied at 1
AU, on the equatorial plane,
\begin{equation}
\label{lambda}
\lambda\simeq\frac{B_{T,1\rm{AU}}}{B_*}\frac{M_{1\rm{AU}}^2}{R_{1\rm{AU
}}} \,.
\end{equation}

By definition, we also have
\begin{equation}
\label{niu} \nu=\sqrt{\frac{2GM}{r_* V_*^2}} \,.
\end{equation}

\begin{table}
\centering \caption{Input and output data for the critical wind
solution calculated using model B, \emph{hybrid 1} and
\emph{hybrid 2}.}\label{table_total}
\begin{center}
\begin{tabular}{c c c c c}
  \hline\\
     &  \small{\emph{input param.}} & \small{\emph{model B}} & \small{\emph{hybrid 1}}  & \small{\emph{hybrid 2}}\\
  \\
  \hline\\
    & $\delta$ & $\;\;1.95$ & $\;\;\;2.90$ & $\;\;\;0.49$\\
    & $\epsilon$ & $\;\;5.64$ & $\;\;\;5.64$ & $\;\;\;5.64$\\
    & $\mu$ & $-\;0.682$ & $-\,0.406$ & $-\,0.029$\\
    & $\kappa$ & $-$ & $\;\;0.20$ & $\;\;\;\;\;0.0123$\\
\\
  \hline\\
   At  & $\eta=V_{1\rm{AU}}/V_*$ & 2.15 & 2.80& 1.90\\
   1 AU  & $\lambda$ & 0.1662 & 0.2468 & 0.1383\\
    & $\nu$ & 1.462 & 0.8872& 0.3767\\
    & $V_{1\rm{AU}}$ (km/s) & 775 & 775 &775\\
    & $n_{1\rm{AU}}$ (cm$^{-3}$) & 2.48 & 2.48 & 2.48\\
    & $B_{1\rm{\small{AU}}}$ ($\mu$G) & 2.432 & 9.81& 30.4\\
    & $B_{\rm{T},1\rm{AU}}$ ($\mu$G) & 29.5 & 29.5 & 29.5 \\
  \\
  \hline
\\
    & $\ \small{\emph{output param.}} $ & & &\\
  \\
  \hline\\
   At & $\eta=V_{1\rm{AU}}/V_* $ & 2.151 & 2.193&1.908 \\
   1 AU & $R_{1\rm{AU}}$ & 156.9 & 137.0 & 13.37\\
   \\
   & $V_{1\rm{AU}}$ (km/s) & 775.4 & 607.6& 779.4\\
   & $n_{1\rm{AU}}$ (cm$^{-3}$) & 2.413 & 0.207&2.414\\
   & $B_{1\rm{AU}}$ ($\mu$G) & 2.369 & 0.641& 29.8\\
   & $B_{\rm{T},1\rm{AU}}$ ($\mu$G) & 29.45 & 9.64& 29.1\\
   & $T_{1\rm{AU}}$ ($10^5$ K) & 3.037  & 4.246  & 41.88 \\
  \\
 \hline\\
   At the  & $V_{*}$ (km/s) & 360.5 & 277.1 & 360.5\\
   \Alf  & $n_{*}$ ($10^3$ cm$^{-3}$) & 131.3 & 8.095 & 0.8423\\
   radius & $B_{*}$ ($10^4$ $\mu$G) & 5.989 & 1.143 & 0.5435\\
   & $B_{\rm{T}*}$ ($10^{3}$ $\mu$G) & 9.953 & 1.360 & 0.278 \\
   & $T_{*}$ ($10^{6}$ K) & 9.892 & 3.902 & 3.597\\
\\
\hline
\end{tabular}
\end{center}
\end{table}

\begin{figure}
\centering
   \includegraphics[width=260 pt]{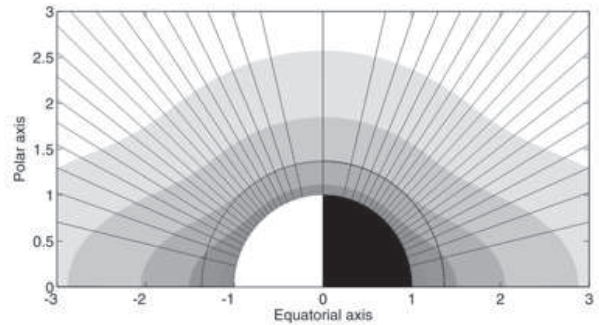}\\
  \caption{Poloidal fieldlines and density contours
of the solution of Table \ref{table_total} for model B. Distances
are given in solar radii. The solid circle line indicates the \Alf
singularity and the fast magnetosonic separatrix which are almost
coincident.}\label{geometry_LPT}
\end{figure}

\begin{figure*}
\centering
    \includegraphics[width=460 pt, height=200 pt]{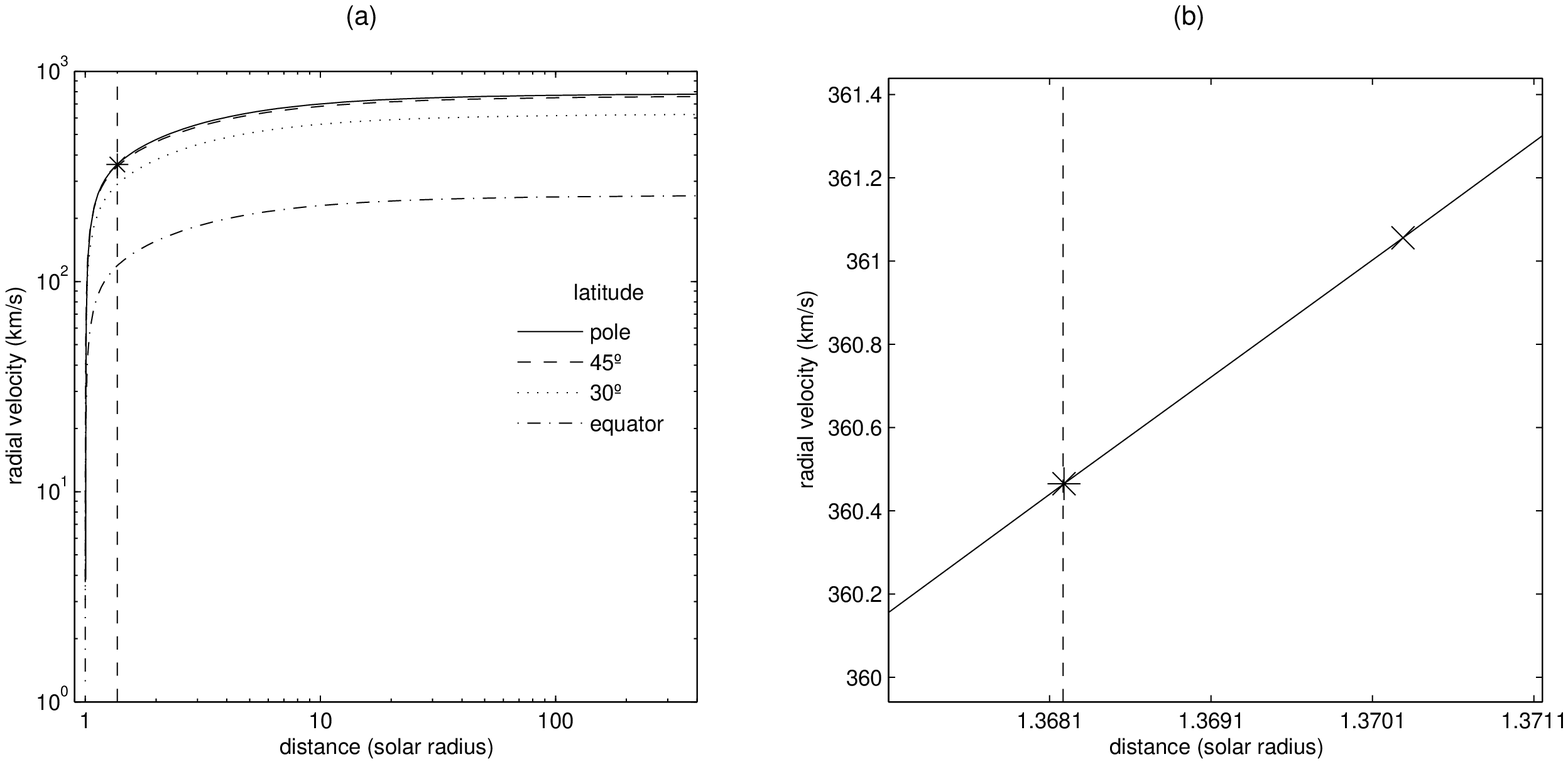}\\
\caption{Panel (a) polar radial velocity as a function of
distance. Panel (b) a zoom view of the singularities: the \Alf and
the fast critical points, represented by $\ast$ and $\ \times$
respectively.}\label{zoom_LPT}
\end{figure*}

Our procedure for finding a critical solution, using model B, that
fulfills all criteria is thus:
\begin{itemize}
\item [-] By fitting \textsc{Ulysses} data at 1 AU obtain the
anisotropy parameters, $\epsilon$, $\delta$ and $\mu$, and the
values $B_{1\rm{AU}}$,  $B_{T,1\rm{AU}}$, $\rho_{1\rm{AU}}$ and
$V_{1\rm{AU}}$; \item [-] calculate $R_\odot$, $\lambda$ and $\nu$
with $f_{1\rm{AU}}=1$ and using Eqs. (\ref{R1au}) to (\ref{niu});
\item [-] with an initial guess of $\eta$, determine the values of
$V_*$ and $B_*$ from Eqs. (\ref{MA1au}) and (\ref{ro_star}); \item
[-] at this stage it is possible to build a critical solution
using the criteria of acceleration continuity at the critical
points; \item [-] this solution will give new values for the solar
surface radius and for the different physical quantities at 1 AU;
\item [-] iterate until the computed values of the solar surface
radius and velocity at 1 AU are close to the fitted ones.
\end{itemize}

Two convergence criteria are inherent to this procedure, the
distance of the \Alf surface above the Sun (or, equivalently, the
value of the magnetic field strength at 1 AU -- see Eq.
\ref{R1au}) and the velocity at 1 AU. Satisfying all the criteria
only by changing $\eta$ is not possible. Therefore, this can only
be achieved by changing, in addition, at least one of the
parameters, thus releasing one of the constraints. Considering
that there are five constraints given from the data fit, exploring
all the parameter space is a formidable task. The set of
parameters concerning the best possible solution is presented in
Table \ref{table_total}. This will be discussed in the following
section.

\subsection{Results}
\begin{figure}
    \includegraphics[width=260 pt]{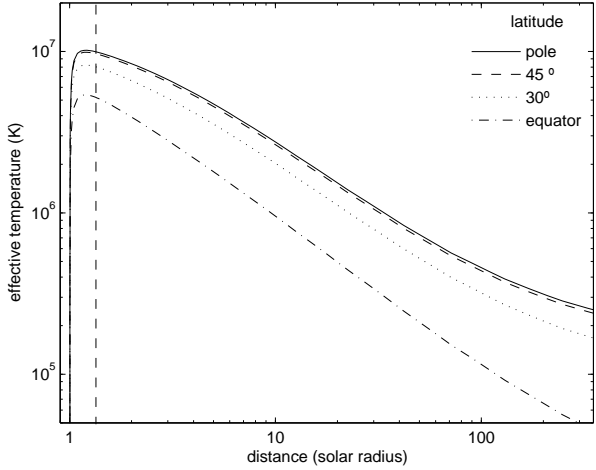}\\
\caption{Effective temperature as a function of distance for
different values of latitude for the model B solution. The
vertical dashed line represents the \Alf radius.}\label{t_LPT}
\end{figure}

%
As can be seen in Table \ref{table_total} we have constructed a
solution where the convergence criteria are quite well fulfilled
with input/output ratios very close to unity for all physical
quantities.  In Fig. \ref{geometry_LPT} we show the fieldlines and
the density contours in the poloidal plane. In Fig.
\ref{zoom_LPT}, where the vertical dashed line represents the \Alf
radius, the acceleration at the critical points is clearly
continuous. Note the presence of two different critical points,
very close to one another in Fig. \ref{zoom_LPT} (b). However,
searching for a converged solution led us to this single set of
parameters by using a value of the radial magnetic field,
$B_{1\rm{AU}}$, of the order of one tenth of the value measured by
{\sc Ulysses}. A similar discrepancy was also found in SLIT05
using the STT99 model instead of LPT01 model. Moreover, the
parameter that evaluates the anisotropy of the radial magnetic
field, $\mu$, has also suffered a shift in its value (compare
Table \ref{table1} to Table \ref{table_total}). As mentioned above
we have tried to change the other input parameters and calculate
their influence on the solution. The best option was to change
those two parameters. Although these trials have been nearly
exhaustive, degenerated sets of the input parameters for the same
critical solution may be possible. In Fig. \ref{t_LPT} we show the
temperature profiles at various latitudes. The higher effective
temperature along the polar axis corresponds to the fast solar
wind. At lower latitudes the lower temperature is related to a
mixing between the fast and slow wind, which also corresponds to
the lower velocities seen in Fig. \ref{zoom_LPT} (a). The
temperature distribution is similar \textbf{to} the one presented
in SLIT05 although its maximum value is slightly better, around
$10\times 10^6$.

\begin{figure}
    \includegraphics[width=260 pt]{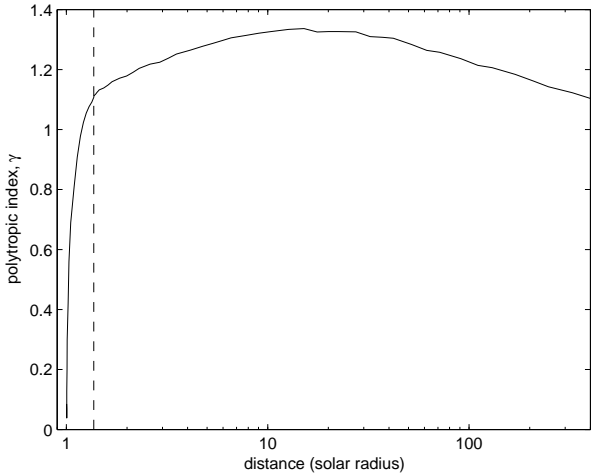}\\
\caption{Polar polytropic index as a function of distance for the
model B solution. The vertical dashed line represents the
Alfv\'{e}n radius.}\label{gama_LPT}
\end{figure}

We have also calculated the effective polytropic index of this
solution. After the temperature peak, it is almost constant,
between $1.1$ and $1.3$ (see Fig. \ref{gama_LPT}). This value is
quite close to the value inferred by Kopp and Holzer
(\citeyear{KoppHolzer76}) in their early polytropic model. Despite
the difference between model A and B in their geometry, neither
can reproduce from the observations the high value of the magnetic
field inferred at 1 AU. The calculations were made such that we
keep the temperature as low as possible and a reasonable value of
the magnetic dipolar field at the base of the corona. We reproduce
successfully the temperature and the velocity profile at 1 AU. It
seems that the geometry does not control the decrease of the
magnetic field but rather the temperature profile. It is even more
surprising that in this LPT01 solution the density at 1 AU remains
at a reasonable level, thus both the velocity and the mass flux at
1 AU correspond to the observed values. It is thus more consistent
to build an hybrid solution combining both models A and B.

\section{Hybrid solutions}
\subsection{Method for a solution}\label{sect_method_hybrid}
For the construction of the hybrid model we consider two different
domains, one from the solar surface to the \Alf radius, and the
other from the \Alf radius towards infinity. In the first domain
we use model A, since the 3D structure is more able to describe
the wind structure near the solar surface. Although the
latitudinal functions of model A are not very versatile it does
provide a full 3D description of the flaring. In the domain
further out, where the fieldlines are almost radial in the
poloidal plane, we can use model B and take advantage of its
flexibility in fitting very steep variations of the physical
quantities with latitude. The border between these two different
domains was arbitrarily set at the \Alf surface.

The major drawback of this construction is that we cannot
guarantee continuity of all physical quantities everywhere except
along the polar axis. Generating the critical solution with model
A means that it must cross both slow magnetosonic separatrix
critical point and Alfv\'{e}nic
 singular point. In addition, the critical solution
with model B has to cross the \Alf point and a fast separatrix
critical point. Thus, a new feature of this hybrid model compared
to our previous solutions (SLIT05 and Sect. \ref{build_B}) is the
crossing of the three usual MHD critical points. Such a situation
was present only in the over-pressured solutions presented in
STT04. Since for the solar wind the fast point is close to the
\Alf one, this was one more argument to construct the hybrid
solution starting precisely at this Alfv\'{e}nic transition.
Moreover, in order to match the two solutions we must search for
continuity of the physical quantities at the boundary as much as
possible. This means that at the \Alf surface we ask for
continuity of the density, pressure, velocity and magnetic field
plus continuity of the acceleration and fieldline geometry.
Strictly speaking, this can only be done along the polar axis
because the latitudinal dependences of the physical quantities are
not identical in both models. Mathematically, we have
\begin{equation}\label{transition_constrains1}
{\rho^{model A}_*}={\rho^{model B}_*}
\end{equation}
\begin{equation}\label{transition_constrains2}
\Pi_*^{model A}+C=\Pi_{0,*}^{model B}
\end{equation}
\begin{equation}\label{transition_constrains3}
{V^{model A}_*}={V^{model B}_*}
\end{equation}
\begin{equation}\label{transition_constrains4}
{B^{model A}_*}={B^{model B}_*}
\end{equation}
\begin{equation}\label{transition_constrains5}
{\frac{dY}{dR}^{model
A}_{\mid_{r_*}}}={\frac{dY}{dR}^{modelB}_{\mid_{r_*}}}
\end{equation}
\begin{equation}\label{transition_constrains6}
{\frac{df}{dR}^{model A}_{\mid_{r_*}}}={\frac{df}{dR}^{model
B}_{\mid_{r_*}}}=0
\end{equation}

The technique used to obtain a full hybrid solution is as follows.
First, the radial velocity at the \Alf point, $V_*$, is determined
by the same procedure as in the previous section. Then, the model
A critical solution has to cross the slow separatrix critical
point and the poloidal fieldlines have to be radial at the \Alf
point, Eq. (\ref{transition_constrains6}). This yields the value
of the velocity slope (i.e. the acceleration) at that transition
point. Similarly, the value of the acceleration at the \Alf point
is determined by crossing the fast separatrix critical point for
the critical solution of model B. In order for this slope of the
velocity to be equal on both sides of the transition point, Eq.
(\ref{transition_constrains5}), the value of $\mu$ had to be
changed from the one derived from \textsc{Ulysses} data. However
this parameter is not very well determined and affects only model
B. In model A its value is fixed to $-1$ by construction and
cannot be fitted. For model A, knowing the slope of the velocity,
Eq. (\ref{transition_constrains5}), and the geometry, Eq.
(\ref{transition_constrains6}), at the \Alf point, the value of
$\Pi_{*}$ is fixed by crossing the \Alf point. Simultaneously, the
value of $\Pi_0$ is fixed by the condition that the pressure is
zero at infinity. This determines the constant $C$ by Eq.
(\ref{transition_constrains2}). Finally, the \Alf distance is
fixed by the magnetic field strength at 1 AU for both models, Eq.
(\ref{R1au}). Thus, the value of $\kappa$ is determined such that
the solution of model A matches the solar surface, $R_\odot$.

\subsection{Results for a hybrid solution - hybrid 1}

\begin{figure}
\centering
   \includegraphics[width=260 pt]{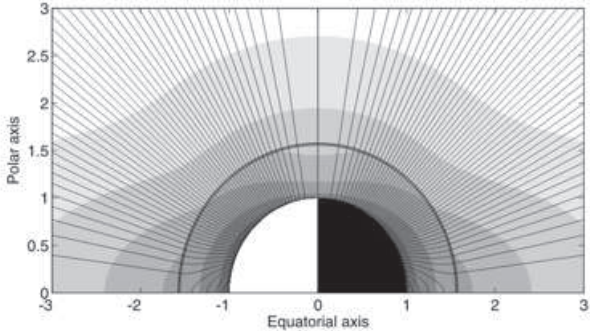}\\
  \caption{Plot of the fieldlines concerning the hybrid
solution and density contours presented in Table
\ref{table_total}, column 3, (\emph{hybrid 1}). Distances are
given in solar radii. The black circles corresponding to the slow
magnetosonic critical separatrix, the \Alf singularity and the
fast magnetosonic critical separatrix, which almost coincide with
each other.} \label{geometry_STT_1}
\end{figure}

\begin{figure*}
\centering
    \includegraphics[width=510 pt, height=210 pt]{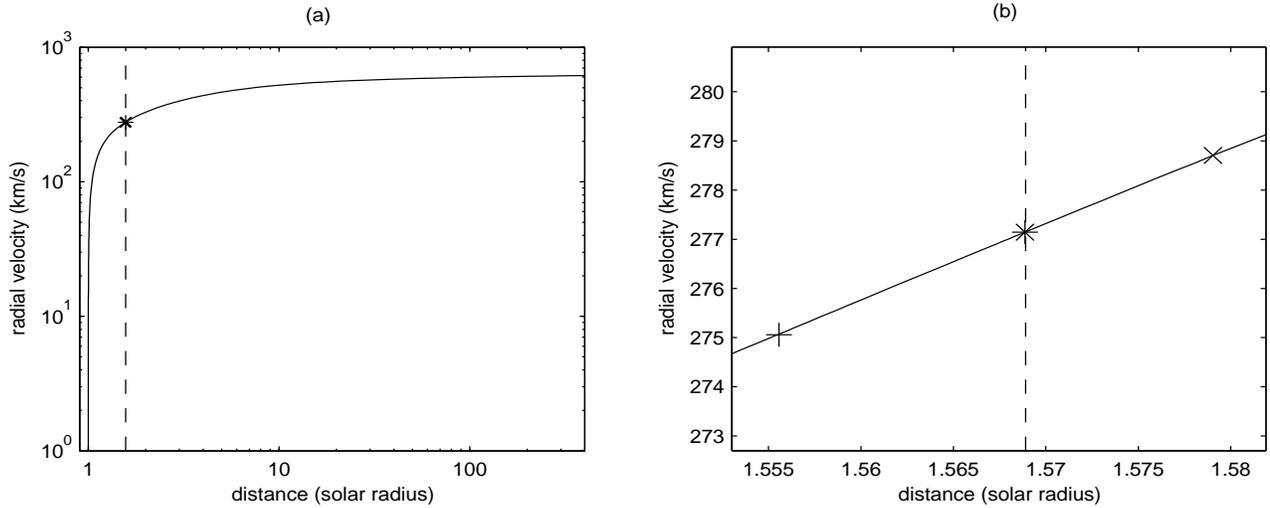}\\
\caption{Panel (a) the polar radial velocity as a function of
distance for solution hybrid 1. The vertical dashed line is at the
transition point (\Alf point). Panel (b), a zoom view of the three
critical points is displayed. The slow magnetosonic, the \Alf and
the fast magnetosonic critical points are labelled with $+$,
$\ast$ and $\times$, respectively.}\label{zoom_STT_1}
\end{figure*}

Table \ref{table_total} shows the input and output values for the
most important physical quantities regarding the first hybrid
solution obtained using the technique presented in the previous
section. In this case convergence between the input and output
parameters is less satisfying than in Sect. \ref{build_B}. This
solution is hereafter referred to solution \emph{hybrid 1}. Figure
\ref{geometry_STT_1} shows the geometry of the fieldlines for this
solution. It clearly shows that beyond the \Alf point the
fieldlines become radial in the poloidal plane and that the
flaring zone near the base of the wind is well defined. Thus
conversely to SLIT05 where the dead zone was too extended, we have
a more realistic geometry. Figure \ref{zoom_STT_1} shows the polar
radial velocity where the vertical dashed line represents the \Alf
radius, which corresponds to the border between application of
models A and B.

\begin{figure}
    \includegraphics[width=240 pt]{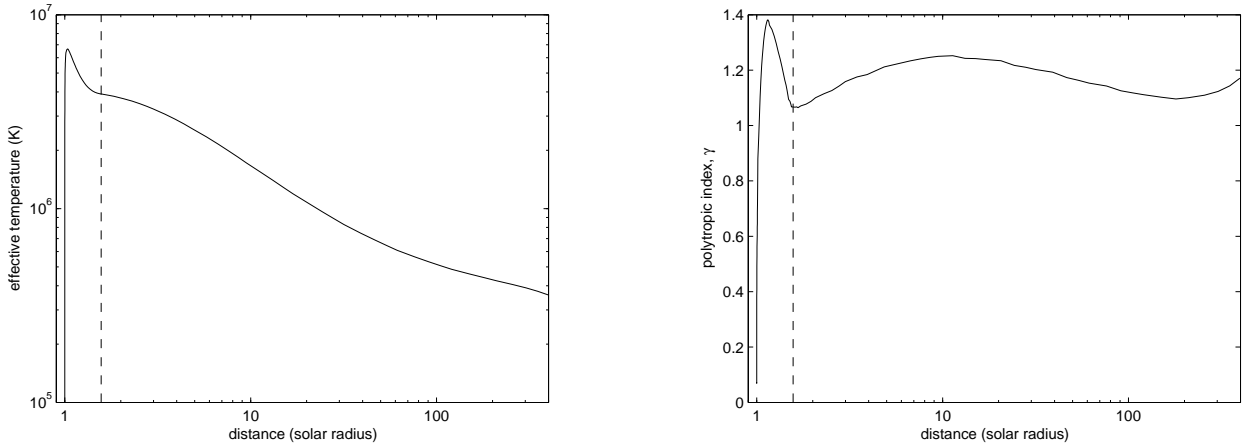}\\
\caption{Effective temperature as a function of distance for the
polar axis, for solution \emph{hybrid 1}. The vertical dashed line
represents the transition point between the use of model A and
B.}\label{t_STT_1}
\end{figure}

\begin{figure}
    \includegraphics[width=240 pt]{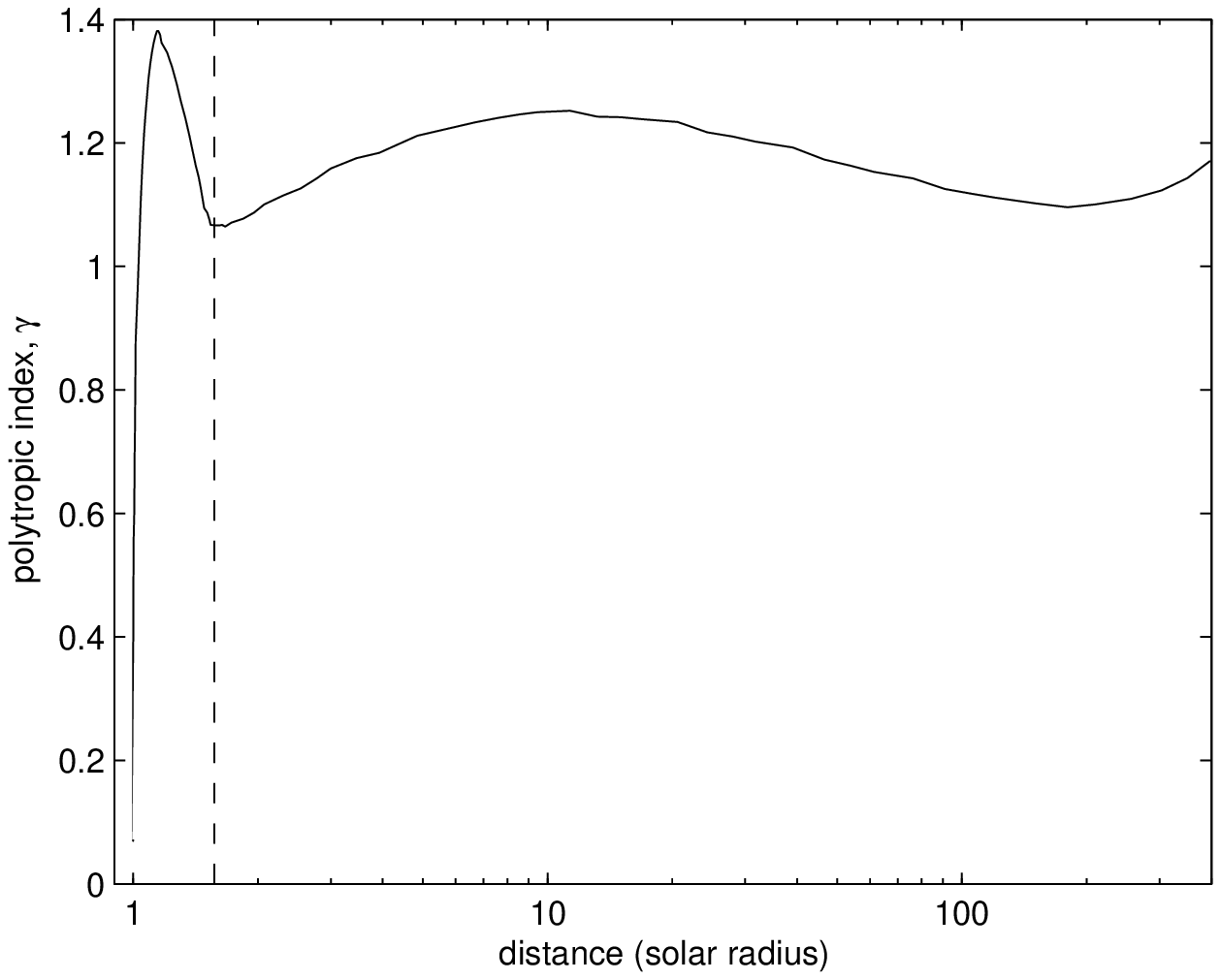}\\
\caption{Radial profile of the polytropic index for solution
\emph{hybrid 1}. The vertical dashed line represents the
transition point between the use of model A and
B.}\label{gama_hybrid_1}
\end{figure}

The presence of three critical points characterizes a different
topology for this kind of wind solution (similar cases were
already discussed in STT04). Figure \ref{t_STT_1} shows the
profile of the temperature for this solution. It is in reasonably
good agreement with observations, namely at 1 AU. The polytropic
index is shown in Fig. \ref{gama_hybrid_1}. It shows, as expected,
that a value around 1.2 is very well adapted to the solar wind,
except in the low corona. However, as we shall discuss in next
section, this first hybrid solution suffers from the same
drawbacks as the previous one, despite its more sophisticated
structure. In addition, the density is too low by one order of
magnitude. Thus, although the temperature profile is low enough to
be in agreement with observations, the mass flux at 1 AU remains
too low. The low effective temperature does not prevent us from
obtaining large velocities but rather from obtaining large
magnetic field at large distances. Thus we reconsidered the values
of some of the parameters, in particular the value of the
latitudinal dependence of the pressure which is not very well
constrained from the observations, to construct another hybrid
solution more fitted to the observed magnetic field.

Another way of analyzing the drawback of this solution is to
examine the convergence of the value $r_\odot/r_*$. It represents
by itself the convergence of the radial magnetic field intensity
at 1 AU, Eq. (\ref{R1au}).The convergence of the values of the
\Alf Mach number and density at this point, Eqs. (\ref{MA1au}) and
(\ref{ro_star}), follows as a consequence. \textsc{Ulysses} data
at 1 AU lead to a very high value of $R_\odot=r_\odot/r_*$ which,
in turn, means that the acceleration of the wind up to the
Alfv\'en speed should take place on a larger scale than the model
predicts. Thus, a more satisfying solution should display a lower
total acceleration from the surface up to the \Alf point. The
fully radial model used (see Sect. \ref{build_B}) is not able to
produce that kind of behaviour. Some degree of flaring is needed
in order to slow down its acceleration. Hence, the only way to
deal with the problem is to decrease the acceleration zone by
decreasing the value of $B_r$ at 1 AU, from Eq. (\ref{R1au}). On
the other hand, in model \emph{hybrid 1}, the high density
gradient near the equator (a consequence of the high values of
$\epsilon$ and $\delta$) provides a poloidal pressure towards the
pole, improving the polar collimation and subsequent acceleration
of the wind. Therefore, the equatorial gradient of the radial
magnetic field must increase ($|\mu|$ must increase), generating a
higher magnetic pressure towards the equator which counterbalances
the previous effect.

This wind solution needs a value for the magnetic field anisotropy
parameter, $\mu$, very different from the one expected. It cannot
easily describe the latitudinal profiles of the physical
quantities from Ulysses data at 1 AU. Its main limitation (besides
the temperature, to which we will refer later) is its discrepancy
on the value of radial magnetic field. Both radial and hybrid
solutions fail completely in reproducing the observed values of
the the magnetic field but the hybrid solution also has a problem
in reproducing the density at 1 AU. This lead us to construct the
hybrid solution in a slightly different way.

\subsection{Results for a fully converged hybrid solution - hybrid 2}
In Table \ref{table_total} we show the input  and output values of
the most important physical quantities, for a second hybrid
solution - hereafter referred to solution \emph{hybrid 2}.
Although we had to release some of the initial values of the
parameters as deduced from SLIT05, this new solution shows a much
better agreement between the initial guesses and the computed
values.

\begin{figure}
\centering
   \includegraphics[width=260 pt]{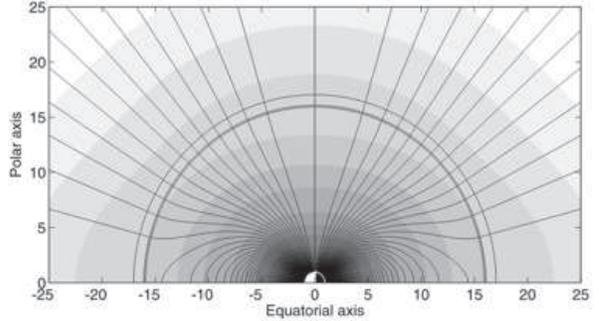}\\
  \caption{Fieldlines and density contours
concerning the hybrid solution presented in Table
\ref{table_total} (\emph{hybrid 2}). Distances are given in solar
radii. The three black circles represent the three surfaces: the
slow magnetosonic separatrix and \Alf singularity, which are
almost coincident, and a fast magnetosonic separatrix.}
\label{geometry_STT_2}
\end{figure}

\begin{figure}
   \includegraphics[width=260 pt]{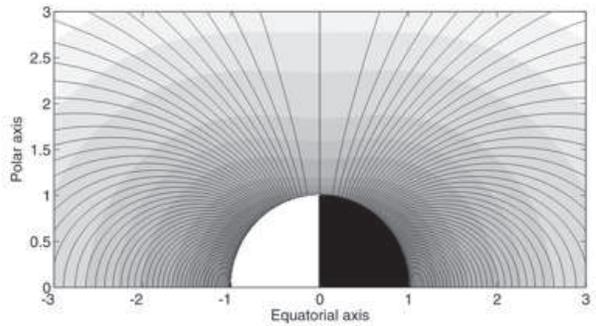}\\
  \caption{\small{Fieldlines and density contours for
solution \emph{hybrid 2}, close to the solar surface. Distances
are given in solar radii.}}\label{zoom_STT_2}
\end{figure}

\begin{figure*}
\centering
   \includegraphics[width=480 pt, height=190 pt]{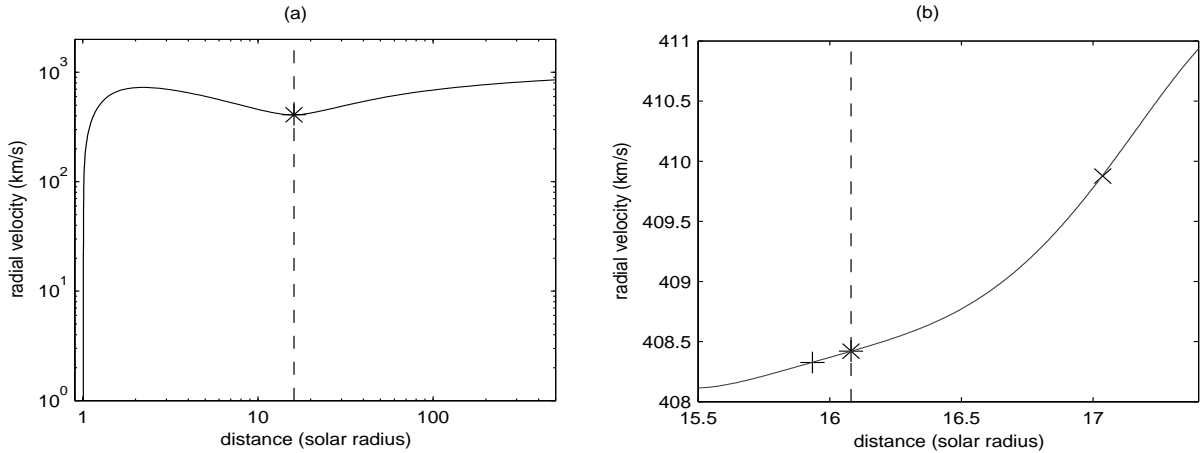}\\
\caption{Panel (a) the polar radial velocity as a function of
distance for solution \emph{hybrid 2}. The vertical dashed line is
at the transition point (\Alf point). Panel (b), a zoom view of
the three critical points: the slow magnetosonic, the \Alf and the
fast magnetosonic critical points, labelled with $+$, $\ast$ and
$\times$, respectively.} \label{zoom_STT}
\end{figure*}

\begin{figure}
    \includegraphics[width=240 pt, height=170 pt]{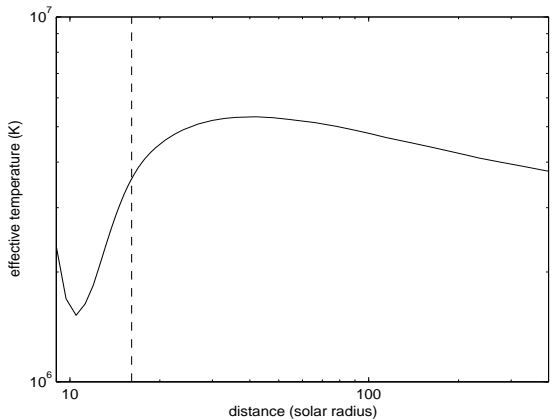}\\
\caption{Effective temperature as a function of distance for the
polar axis, for solution \emph{hybrid 2}. The vertical dashed line
represents the transition point between the use of model A and B.}
\label{t_STT_2}
\end{figure}

\begin{figure}
    \includegraphics[width=240 pt,height=170 pt]{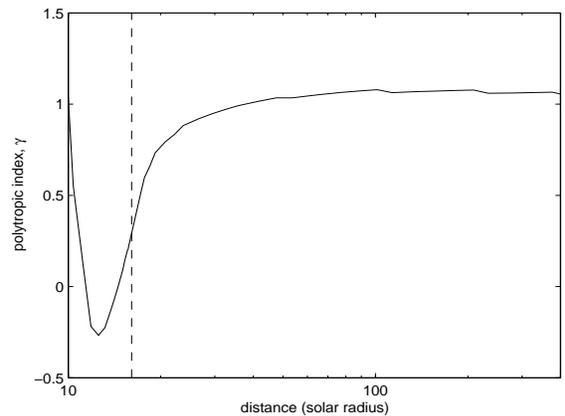}\\
\caption{Radial profile of the polytropic index for solution
\emph{hybrid 2}. The vertical dashed line represents the
transition point between the use of model A and B.}
\label{gama_hybrid_2}
\end{figure}

\begin{figure*}
\centering
   \includegraphics[width=520 pt, height=334 pt]{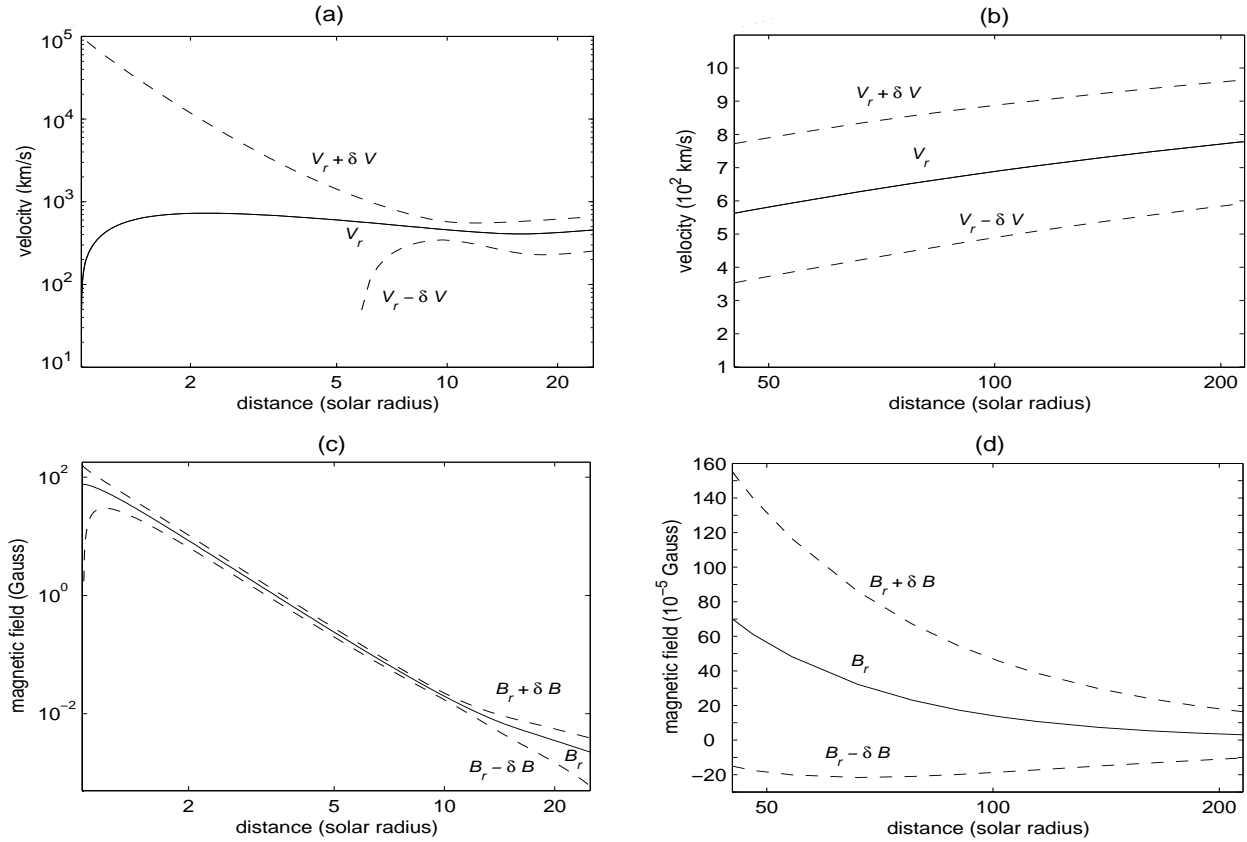}\\
  \caption{Assuming that turbulence
and/or \Alf waves can account for the effective temperature of the
solutions, the corresponding bulk radial velocity plus and minus
the velocity fluctuations. (a) near the solar surface and (b)
close to 1 A.U. The magnetic field fluctuations, assuming \Alf
waves, in (c) near the solar surface and (d) close to 1
A.U.}\label{deltaVB}
\end{figure*}

For this solution, we had  to change  the value of $\kappa$ and of
$\delta$, although in a less dramatic way. Changing the value of
$\kappa$ in the sub-\Alfvenic regime is not a serious problem.
First because, in the super-\Alfvenic region where the value of
$\kappa$ is determined for the {\sc Ulysses} data, we use model B
in which we have no control over the latitudinal dependence of the
pressure. Second, fitting the value of $\kappa$ in this domain is
almost irrelevant as we do not really control the kinetic
temperature (and pressure) which is the real temperature measured
by the spacecraft.  This discrepancy in the parameter can easily
be evaluated by comparing input and output values of the same flow
quantities. Figs. \ref{geometry_STT_2} and \ref{zoom_STT_2} show
the geometry of the fieldlines and how it changes from model A  to
model B. A new feature of this solution can be seen in Fig.
\ref{zoom_STT} (a) - a zone where the radial velocity attains a
local minimum, close to the \Alf point. Figure \ref{t_STT_2}
displays the  effective temperature profile along the polar axis.
This plot clearly shows that the kinetic pressure alone cannot
account for the effective temperature. Figure \ref{gama_hybrid_2}
shows the corresponding polytropic index.
The excess between this temperature and the observed one can be
accounted for, as in SLIT05, assuming extra pressure from \Alf
waves or turbulent or ram pressure (Fig. \ref{deltaVB}). We have
calculated the amplitude of the ram velocity/magnetic field
fluctuations, if the effective temperature we have calculated is
assumed to be the result of turbulence/\Alf waves. For ram
pressure we assume:
\begin{equation}\label{Pram}
    P_{\rm{ram}}=\frac{1}{2}\rho \delta V^2,
\end{equation}
and for the \Alf pressure we take
\begin{equation}\label{Pram}
    P_{\rm{Alfven}}=\frac{\delta B^2}{8\pi},
\end{equation}

These calculations are arbitrary and a mixing of various processes
is probably the source of the extra pressure that accelerate the
fast  wind. However it gives an order of magnitude of the
fluctuations needed. Comparing the various plots a), b), c) and d)
in Fig. \ref{deltaVB}, we conclude that \Alf waves are more
appropriate to explain the acceleration in the sub-Alfv\'{e}nic
part where the magnetic field is dominant. Conversely, turbulence
and fluctuations of the velocity may account for the acceleration
in the super-\Alfvenic region. We arrive at this conclusion only
because the calculated fluctuations of the magnetic field in the
sub-Alfv\'enic region are smaller than the calculated turbulent
velocity field and the reverse holds in the super-Alfv\'enic part.
This is the best way to minimize the amplitude of the fluctuations
in both regions. A mixture of the two components is probably more
realistic but this needs a more detailed model to interpret the
role of turbulence in heating the flow.

\begin{figure}
    \includegraphics[width=260 pt]{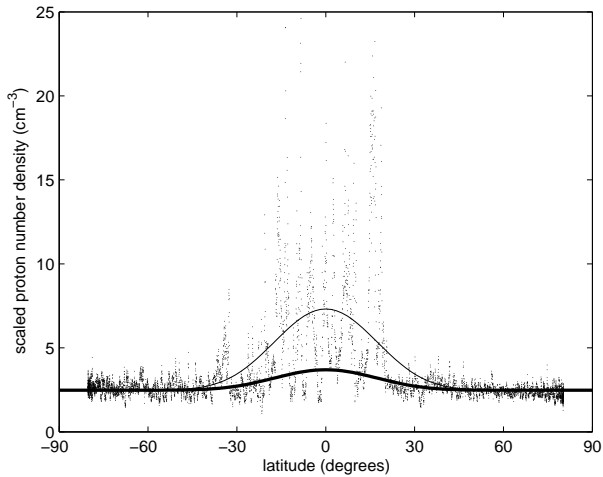}\\
\caption{Scaled proton density with latitude. The points represent
hourly averaged data from the \textsc{Ulysses} Swoops Ions
experiment. The solid curve corresponds to the fit using model B
and the bold solid curve to the solution generated by solution
\emph{hybrid 2}.}\label{ro_fit}
\end{figure}

\begin{figure}
    \includegraphics[width=260 pt]{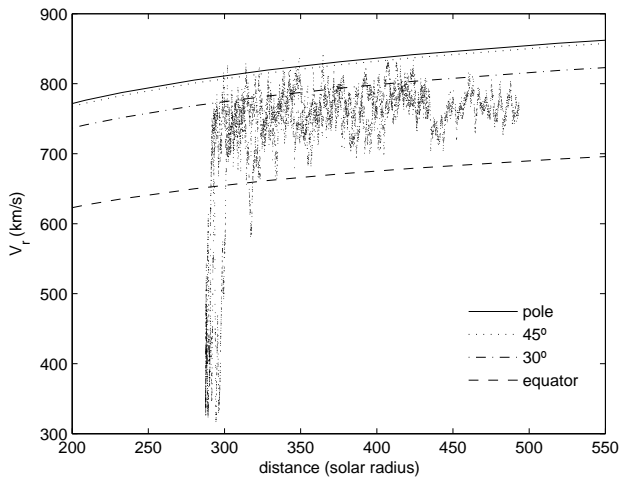}\\
\caption{Radial velocity with distance to the Sun. The points
represent hourly averaged data from the \textsc{Ulysses} Swoops
Ions experiment. The curves correspond to the radial velocity at
different latitudes of the solution generated by solution
\emph{hybrid 2}.}\label{vr_fit}
\end{figure}

This new hybrid solution generates a field geometry that is
continuous at the transition point (it is still not differentiable
and kinks in the field are unavoidable) and shows features
expected for the solar wind (Figs. \ref{geometry_STT_1} and
\ref{geometry_STT_2}). It is also capable of reproducing almost
all \textsc{Ulysses} data at 1 AU. Despite the slight difference
between the values of the anisotropy parameters when calculated by
fitting the  data (Table \ref{table1}) and the one from the
critical solution itself (Table \ref{table_total}), most of the
limitations of the previous solutions have been solved. Such a
discrepancy should not be very important since the new values can
easily be fitted (with some degree of accordance) to the observed
data (see for instance the fit of the density in Fig. \ref{ro_fit}
and of the radial velocity in Fig. \ref{vr_fit}). The ratio
between the input and the output values for the most important
physical quantities is very close to unity and all the continuity
criteria are satisfied. Nevertheless, $\kappa$, $\delta$ and $\mu$
have values departing from the expected ones. The value of $\mu$
is the result of the transition conditions stated in Eqs.
(\ref{transition_constrains1}) to (\ref{transition_constrains6})
and has been calculated accordingly.


The values of the magnetic field intensity at 1 AU constrain the
value of $R_\odot$, and therefore the length of the wind
acceleration zone (or the dead zone). Consequently, the physics
controlling the hybrid model forced us to adapt it in order to
obtain the required feature. Reminding that $\kappa$ and $\delta$
characterize the anisotropy of the pressure and density,
decreasing both parameters will lead to a decrease of the pressure
gradient towards the pole, which enables the wind to accelerate
more slowly (from the solar surface to the \Alf point). As a
consequence of acceleration continuity at the transition point,
$|\mu|$ also diminishes which means that the magnetic pressure
gradient towards the equator, outside the \Alf sphere (in the
fully radial zone of the model), also decreases. For the dynamics
of the radial part of the hybrid model, the wind velocity is
expected to be higher in order to satisfy the values at 1 AU and
so it needs to accelerate the wind. This leads to a decrease in
the magnetic pressure towards the equator and thus a decrease of
$\mu$.

Of course, there is a price to pay to fit all data at 1 AU. Thus,
this hybrid model has two major drawbacks. The physical quantities
are discontinuous at the transition radius except for the polar
ones. This is a natural consequence of the analytical expressions
that describe the flow, Eqs. (\ref{STTVr}) to
(\ref{LPT_model_end})  and the temperature behaviour. The first
one is solved only for values of $\epsilon=1$. The second and more
serious drawback is the very high effective temperature. This can
be explained only if we calculate the heat flux using a reasonable
kinetic theory \citep{Zouganelisetal05} together with solving a
full energy equation. This amounts to invoking a non thermal
heating term, a difficult task that we postpone for future work.
In Fig. \ref{t_STT_2}, we see how the energy equation can be
essential. The absence of the an abrupt increase of the
temperature very close to the surface  in other models, such as
the one presented in SLIT05 and the one presented in Sect.
\ref{build_B} of the present work might be explained by solar
surface being much closer to the \Alf radius and therefore the
problems had not emerged yet. Nevertheless, high temperatures are
reached (for an overall behaviour of the wind solution) as a
consequence of high values of the magnetic field at 1 AU and not
necessarily high values of velocity as one may expect intuitively.

\section{Conclusions}
From the constrained parameters obtained after fitting
\textsc{Ulysses} data (SLIT05) we were able to build different
critical solutions for the solar wind. The first was obtained
using a purely radial field (model B). The remaining two solutions
where constructed as hybrid ones incorporating an inner region
where model A (with flaring streamlines) was used and an outer one
with model B. Thus, we combine the advantages of model B of
reproducing highly adaptable functions of latitude and the
advantages of model A of ensuring adequate flaring of the
fieldlines to get a more realistic geometry of the overall solar
wind. These two distinct models were coupled using well defined
domains for each one and a suitable transition zone, the \Alf
radius. Both model B (used by itself) and the hybrid model (A and
B coupled) were used to generate a critical wind solution that
could replicate the measured values of some important physical
quantities, such as the radial velocity, the radial and toroidal
magnetic fields and the density, at 1 AU measured by {\sc Ulysses}
at solar minimum.

In the hybrid model, we could not avoid some discontinuity out of
the polar axis. This may be solved in the future by using the
present solutions to start more realistic MHD simulations. For the
first model with helicoidal streamlines, the computed velocity and
density values at 1 AU are in good agreement with \textsc{Ulysses}
data. However, we were clearly not able to reproduce the values of
the radial magnetic field. This is also true for the first hybrid
model we presented (\emph{hybrid 1}), which in addition failed to
reproduce the density at  1 AU by one order of magnitude. This can
be understood because we tried to minimize the temperature along
the flow. The second hybrid model is able to reproduce the values
of all physical quantities at 1 AU except for the temperature.
This is too high even though we can invoke non thermal processes
to explain the excess of effective pressure. It also provided a
solution with a smooth geometry where the fieldlines became purely
radial after the \Alf radius. Some concessions were made in the
values of the parameters that rule the dynamics of both models.
They are slightly different from the ones calculated in the method
presented in SLIT05. Such differences can be explained by the
physics that describes the dynamics of the flow. Even though the
high values of the temperature at 1 AU can be explained (SLIT05),
its behaviour close to the solar surface suggests the
implementation of an energy equation. Since this is a formidable
task, we postpone it for future work. However, we face the same
difficulties as any MHD simulation. The main advantage of the
present solutions is its simplicity. Moreover, they are exact
solutions of the ideal MHD equations.

\section*{Acknowledgments}
The authors acknowledge the French and Portuguese Foreign Offices
for their support through the bilateral PESSOA programme and
the SWOOPS instrument team.  The present
work was supported in part by the European CommunityÕs
Marie Curie Actions - Human Resource and Mobility within the JETSET (Jet
Simulations, Experiments and Theory) network under contract MRTN-CT-2004
005592. A.A. and J.L also acknowledge support from grant
POCI/CTE-AST/55691/2004 approved by FCT and POCI, with funds
from the European Community programme
FEDER.

\bibliography{bibdata}
\bibliographystyle{NBaa}
\end{document}